\documentstyle[11pt]{article}
 
\def\pa{\partial}

 \def\G{\Gamma}
\def\a{\alpha}
\def\b{\beta}
\def\d{\delta} 
\def\e{\epsilon}

\def\l{\lambda} 
\def\m{\mu}
\def\n{\nu}

\def\r{\rho}
\def\s{\sigma} \def\S{\Sigma}

\renewcommand{\baselinestretch}{2.0}

\begin{document}

\renewcommand{\baselinestretch}{1}
\small
\normalsize
\noindent \hspace*{5in}Brandeis BRX--343\\
\hspace*{5in}SISSA 14/93/EP

\renewcommand{\baselinestretch}{2}
\small
\normalsize

\begin{center}
{\bf Geometric Classification of Conformal Anomalies\\
in Arbitrary Dimensions}

\vspace{.15in}

\renewcommand{\baselinestretch}{1}
\small
\normalsize
S. Deser\\
The Martin Fisher School of Physics\\
Brandeis University\\
Waltham, MA 02254, USA

\vspace{.1in}

A. Schwimmer\footnote{On leave from the Weizmann Institute,
Rehovot, Israel} \\
SISSA and INFN Trieste\\
Trieste 34013, Italy
\end{center}

\renewcommand{\baselinestretch}{2}
\small
\normalsize

\noindent{\bf Abstract}

We give a complete geometric description of conformal anomalies in
arbitrary, (necessarily even)
dimension.  They fall into two distinct
classes: the first, based on Weyl invariants that vanish at integer
dimensions,
arises from finite -- and hence scale-free -- contributions to the effective
gravitational action through a
mechanism analogous to that of the (gauge field) chiral anomaly.  Like
the latter, it is
unique and
proportional to a topological term, the Euler density of the
dimension, thereby preserving scale invariance.  The contributions of
the second class, requiring introduction of a scale through
regularization, are correlated to all local conformal scalar polynomials
involving powers of the Weyl tensor and its derivatives; their number
increases rapidly with dimension.  Explicit illustrations in dimensions
2, 4 and 6 are provided.

\vspace{.2in}

Conformal (Weyl) anomalies have a long history (see for example [1--3])
and
are still
being studied (a partial list is in \cite{004}), not least because of their
many applications
({\it e.g.}, [5--9]).
However, their underlying geometric basis has not been systematized
nor extended to general dimensions.  It has also not been appreciated
that there are in fact two distinct types of anomalies (one of which is
very similar to the chiral anomaly!), with quite different
geometric and physical antecedents,
reflecting the distinction
between Weyl and constant scale transformations (dilatations).
We intend here to fill these gaps.

We recall that,
in (even) integer dimensions, the effective gravitational action
generated
by a conformally invariant matter system ({\it e.g.}, massless spinors
or improved
massless scalars, d=4 photons)  contains contributions that cannot
simultaneously preserve diffeomorphism and
Weyl symmetries.
For free matter, the anomaly is manifested in the clash between
conservation
and tracelessness of stress tensor correlators.  We will of course opt to
retain conservation {\it i.e.}, diffeomorphism invariance; a dilatation
then becomes equivalent to a scale (constant Weyl) transformation.
Defining
the infinitesimal change by
\begin{equation}
\d \, g_{\mu\nu} = 2\phi (x) \, g_{\mu\nu}\;,
\end{equation}
and the effective action obtained by integrating out
the matter field by $W[g_{\mu\nu}]$, the conformal anomaly is given
by
\begin{equation}
{\cal A} (g_{\mu\nu}) \equiv \d W/\d \phi (x) \; .
\end{equation}
Now if the effective action is dilatation invariant, {\it i.e.},
contains no scale $\m$, then the anomaly must have vanishing
integral,
\begin{equation}
\d W /\d \ln \m^{2} = \int \, d^d x \, {\cal A} = 0\; .
\end{equation}
We call this a type A anomaly; the scalar density ${\cal A}$ must
therefore be related to a topological invariant, and the only available
parity-even candidate is the Euler density, the analogue of the chiral
case ${\cal A} = F^*F$.  If, on the other hand, $W$ does contain a
scale, then the corresponding anomaly must reflect this,
\begin{equation}
\d W /\d \ln \m^{2} = \int \, d^d x \, {\cal A} \neq 0
\end{equation}
even though ${\cal A}$ itself does not depend on $\m$: this is our type
B anomaly.  [There is also a third, trivial, type of anomaly which is
local, {\it e.g.}, ${\cal A} = \Box
R\: \sqrt{-g}$ in $d = 4$,
and can be removed by a local counterterm.   We will not be concerned
with these.]
The basic tool that will enable us to secure the desired geometric basis
for both classes of anomalies is dimensional regularization (applicable
here because there is no chiral problem; we do not consider
gravitational chiral anomalies): in non-integer dimension $d
= 2(n + \e )$, both symmetries can be preserved simultaneously since
everything is convergent.  This will enable us to enumerate
geometrically all
relevant invariant terms in the effective action.  In this scheme, the
anomaly is of course produced by the $1/\e$ poles that correspond to
the ultraviolet divergences.  We distinguish two possible situations:  If
the coefficient of $1/\e$ itself vanishes as $\e$, the
effective action is
finite, but ambiguous; we then define the limit in such a way that
diffeomorphism invariance is preserved, at the price of a Weyl anomaly.
Since no scale is required to achieve a finite result, this is type A.
We will study the kinematical identities that lead to this situation, and
show that the anomaly is indeed unique, namely the Euler density as
discussed above.  If, on the other hand, the coefficient of $1/\e$ is non-
vanishing, then there is a logarithmic divergence and a scale-dependent
counterterm is required in order to yield a finite result.  In this type B
case, the number of
possible terms increases with dimension,  reflecting
the various conformally invariant combinations of powers of the Weyl
tensor and its derivatives in dimension $d$.

      Our curvature conventions for the background (torsionless)
geometry are
$R^\m_{~\n\a\b} \sim  + \pa_\a \G^\m_{\n\b}, \; R_{\n\b} \equiv
R^\m_{~\n\m\b}$.  The Weyl tensor, which shares the algebraic
symmetries of the Riemann tensor, is
\begin{eqnarray}
C^\m_{~\n\a\b} & = & R^\m_{~\n\a\b} - [\d^\m_\a \:
\tilde{R}_{\n\b}
+ {\rm symm}] \nonumber \\
\tilde{R}_{\n\b} & \equiv & (d -2)^{-1} [R_{\n\b} - 1/2 (d-1) g_{\n\b}
\,R] \; .
\end{eqnarray}
We shall frequently only work to leading order in powers of
$h_{\mu\nu} \equiv
g_{\mu\nu} - \eta_{\mu\nu}$  about flat space.

Let us begin with the simplest, $d = 2$, case to illustrate
our approach in an explicit perturbative framework.
A straightforward one-loop calculation yields the correlator of two
conserved traceless stress tensors in $d = 2 (1+\e)$,
\begin{eqnarray}
\langle T_{\mu\nu}(q)T_{\rho\s} (-q) \rangle & = &
c(d) \G (2-d/2) (q^{2})^{-1 + \e} \, \e^{-1}
[ \textstyle{\frac{d-1}{2}} (P_{\m\rho} P_{\n\s} + P_{\m\s}
P_{\rho\nu}
) - P_{\mu\nu}P_{\rho\s}
] \; ,
\nonumber \\
P_{\mu\nu} & \equiv & \eta_{\mu\nu} q^2-q_\m q_\n
\end{eqnarray}
where $q$ is the momentum carried, the constant $c(d)$ depends on the
Lorentz
structure of the matter field but is  finite  for any
$d>1$, and
$P_{\mu\nu}$ is the transverse projector, whose trace is $(d -1)q^{2}$.
The kinematics of the numerator is uniquely fixed by conservation and
($d$-dimensional)
tracelessness.   While the coefficient of $q_\m q_\n q_\rho q_\s$ in (6)
is
$(d-2)$, making this term ultraviolet finite as required by power
counting,
finiteness in $d =2$ of the remaining structure is achieved, not
through an explicit $(d-2)$ factor, but by the kinematical identity
$P_{\mu\nu} =
\tilde{q}_\m\tilde{q}_\n \, , \;
\tilde {q}_\m \equiv \e_{\mu\nu}q^\n$
valid
only in $d=2$.  Hence
the whole bracket in (6) vanishes there, so that
the $\e
\rightarrow 0$ limit of (6) is ambiguous.  We shall define it by taking
the projector
$P_{\mu\nu}$ to have its $d=2$ form $\tilde{q}_\m\tilde{q}_\n$ (but
keep the explicit ($d-1$)) thereby respecting the conservation Ward
identities (since
$q\cdot \tilde{q}\equiv 0$).  With this prescription, (6) limits to
\begin{equation}
\langle T_{\mu\nu} \, T_{\rho\s} \rangle_{d=2} =
c(2)q^{-2} \, P_{\mu\nu} \, P_{\rho\s} \; ,
\end{equation}
which violates the tracelessness condition, signalling a Weyl
anomaly.\footnote{This violation mechanism becomes even more
transparent if we consider the single $<T_{\m\n} >$ conserved and
traceless structure, $\sim G_{\mu\nu} + \frac{1}{2} \, (\frac{d-2}{d-1})
P_{\mu\nu} \, R/\Box$, in which the first term vanishes at $d=2$,
while the second has the explicit $(d-2)$ factor.  We recall, incidentally,
that in dimensional regularization all tensorial quantities are to be
evaluated at their integer value; for example, it would be incorrect here
to ``continue" $G_{\mu\nu}$ so as to make the above $< T_{\mu\nu}>$
vanish.}
The above result easily translates into a geometric effective action,
since (6) is the leading term in the expansion of
\begin{eqnarray}
W[g_{\mu\nu} ] & = & c(d) \, \G (1-\e ) \, 1/\e \, \int \, d^dx \,
\sqrt{-g} \left[ C^{\mu\nu}_{[\rho \s} \, \Box^{\e -1} \,
C^{\rho\s}_{\m\n ]}
- R^{\mu\nu}_{[\r\s} \Box^{\e -1} R^{\rho\s}_{\mu\nu ]}\right]
\nonumber \\
& = & c(d) \, \G (1-\e ) \, 1/\e  \int d^dx \, \sqrt{-g}
\left[ (d-1) R_{\a\b} \, \Box^{\e -1} R^{\a\b} -
\frac{d}{4} \, R \, \Box^{\e -1} R \right] \; ,
\end{eqnarray}
where
square brackets denote complete antisymmetrization and
fractional powers of the covariant Laplacian are defined by say
\begin{equation}
\Box^\a = \int^{\infty}_0 \, dt \, t^\a /(t-\Box )\; .
\end{equation}
The limit of (8) is taken by using the Ricci tensors defined in $d=2$, {\it
i.e.}, fulfilling the $d=2$ relation
\begin{equation}
G_{\mu\nu} \equiv
R_{\mu\nu} - \textstyle{\frac{1}{2}} \, g_{\mu\nu} \, R = 0
\Longrightarrow
R_{\mu\nu} R^{\mu\nu} = d/4 \, R^{2} \; .
\end{equation}
Our $W[g]$ then reproduces\footnote{One also finds (11) by
multiplying (7) by $h_{\mu\nu}h_{\rho\s}$ and integrating, since
$P^{\mu\nu}h_{\mu\nu}$ is just the leading order of $R$.  There are
no higher order corrections to (11) as it is the only covariant, scale
invariant that is kinematically permitted; note that in $d=2$, $\d_\phi
\Box^{-1} = 2\phi \times \Box^{-1}$ exactly.}
the famous result \cite{005}
\begin{equation}
W_{d=2} = c(2)/2 \int \, d^{2}x \, \sqrt{-g} \, R \, \Box^{-1}\, R
\end{equation}
with the corresponding Weyl anomaly
\begin{equation}
{\cal A}_{d=2} = \d W/\d \phi (x) = c(2) \, \sqrt{-g} \, R \; ,
\end{equation}
proportional to the Euler density $E_2$ in $d=2$. In geometric
language, this
type A anomaly (there are no type B anomalies in $d=2$, as
we shall see that
no suitable
conformal invariants are available there) was a consequence of the
vanishing in $d=2$ of the bracketed expression in the first part of (8),
namely of $C^{2}_{\m\n\a\rho} - E_4$; at $d=2$,
$C_{\mu\nu\a\rho}$
itself vanishes, as does the Euler density $E_4$.

Generalization of the above example to arbitrary dimension requires
knowledge of the Weyl--invariant polynomials in the curvature that
vanish in each higher integer dimension, a question that has been
solved in all generality \cite{010}.
These polynomials arise from the obvious fact that the total
antisymmetrization
of any expression involving 2n indices vanishes
for any integer $d = 2m < 2n$; in particular, consider
a quantity $A^{\mu\nu}_{\rho\s}$ with the algebraic symmetries of
the Riemann tensor, for which
\begin{equation}
\sqrt{-g} \, A^{\m_1\n_1}_{[\m_1\n_1} \ldots
A^{\m_n\n_n}_{\m_n\n_n ]} \equiv 0 \;,  \hspace{.3in} m < n \; .
\end{equation}
{}From (13), there follow two independent identities of relevance to us.
The first is that in which $A=C$ itself.  The second is one in which $A$
is replaced
by $C^{\mu\nu}_{\rho\s} + (\d^\m_\rho K^\n_\s +$ symm), where
$K$ is traceless, in which case the coefficient of the term linear in $K$
in (13) is
\begin{equation}
(d -2 n+1) \sqrt{-g} \, \left(
C^{\m_1\n_1}_{[\m_1\n_1} \ldots C^{\m_{n-1}\a}_{\m_{n-1}\b ]} -
d^{-1} \, \d^\a_\b \, {\rm tr} (C \ldots C) \right) \equiv 0 \; ,
\hspace{.3in} m < n
\end{equation}
in an obvious notation.
The odd prefactor in (14) is responsible, in this language, for the
absence of odd-dimensional anomalies.
[If $K$ is a pure trace, one obtains the less
relevant statement that $tr\,C^n =0$ in lower dimensions.]  If  $A$ is
replaced
by the Riemann tensor itself, then the left side of (13) is just the Euler
density $E_{2n}$ in $d=2n$, expressing the well known fact that
$E_{2n}$
vanishes in all lower integer dimensions; $E_{2n}$ is also a total
derivative
both at $2n$, and -- to lowest order in $h$ -- in any (also non-integer)
dimension.  Since
$E_{2n}$ is obviously a linear combination of (13) with $A=C$ and (14)
times $R^\b_\a$, there is only one relevant contribution of the form
(13) that vanishes for lower integer $d$ and is a conformal invariant to
lowest order in $h$.  Calculations are formally simplified by choosing
\begin{equation}
I_n =
\sqrt{-g} \,
C^{\m_1\n_1}_{[\m_1\n_1} \ldots C^{\m_n\n_n}_{\m_n\n_n ]}  -
E_{2n} \; ,
\end{equation}
which removes the explicit highest powers of the Riemann tensor;
$I_n$ Weyl--transforms homogeneously, {\it i.e.}, only due to the
undifferentiated metrics, $\d I_n = (d-2n)I_n\phi$ (recalling that it is
the
tensor $C^\m_{\n\a\b}$ that is untransformed).  By power counting,
then, the conformally invariant contribution to the effective action will
have the (leading, see below) form
\begin{equation}
W_n = (d-2n + 2)^{-1} \int \, d^d x \, I_n \,
\Box^{\frac{d-2n}{2}} + \ldots \; ;
\end{equation}
The undifferentiated inverse metric in the $\Box$ cancels the above
$\d I_n$.  We need
not specify where the $\Box$ factor acts in (16) because the anomaly
is produced in the corner of the phase space of the momenta $q_i$
where all $q^{2}_i =0$, so that the difference between two terms in
which $\Box$ acts on two different factors in $I_n$ will be
non-anomalous.  We need also only calculate in leading $h$ order; the
all-orders completion is uniquely specified by diffeomorphism
invariance (contributions from higher $I_n$ would vanish).  As
explained earlier, $W_n$ produces the type $A$
anomaly in $d = 2n-2$ because of the ``0/0" mechanism that underlies
it, since $I_n$ vanishes at $d = 2n -2$.  The diffeomorphism-preserving
$d \rightarrow 2n -2$
limit of $I_n$ in (16) is
(as in the $d=2$ example) that in which
the special properties of
Riemann tensors defined in the integer dimension $(2n-2)$ are used,
but
otherwise keeping the explicit $d$-dependence in $W_n$.

The $d=4$ anomaly provides a good illustration.  Here
\begin{equation}
I_3 = \sqrt{-g} (C_{\mu\nu}^{\a\b} C_{\a\b}^{\l\s}
C_{\l\sigma}^{\mu\nu}
-4 C^{\mu\nu}_{\a\b} C_{\mu\rho}^{\alpha\sigma}
C^{\b\rho}_{\nu\sigma}) -E_6 \; .
\end{equation}
A tedious calculation leads to the result
\begin{equation}
W_{3,d=4} = \int \, d^4x \, \sqrt{-g} \, \Box^{-1}
\left[ R^2_{\mu\nu\a\b} R + 10 \,
R_{\mu\nu}R^{\n\a}R_\a^\m - 13 \,
R^2_{\mu\nu}R + \textstyle{\frac{41}{18}} \, R^3 +
6 \, R_{\mu\nu\a\b} \, R^{\m\a}R^{\n\b} \right] \; .
\end{equation}
for the leading term of the
corresponding $W_3$.  Taking the $\phi$-variation of (18) is also a
tedious process; it is easiest
to exploit the fact that the contribution to the anomaly can be reached
along any path in phase space \cite{011,012}, by using the symmetric
one along which
all $q^{2}_i = q^{2}$, with $q_i \cdot q_j = -\frac{1}{2}\,  q^{2}$,
$(i \neq j)$, consistent with
the constraint $(\S q_i )^{2} = 0$.  [This approach generalizes to
arbitrary dimension, with $q_i \cdot q_j = -1/n \, q^{2}$.]  In particular,
we
can move the $\Box^{-1}$ freely and replace $D_\m X \; D^\m Y$ by
$-\frac{1}{2} \Box XY$; to leading order, all derivatives
commute.  The result is as expected,
\begin{equation}
\d \, W_{3,d=4} = 9 \int \, d^4x \, \sqrt{-g} \:
\left[ R^{2}_{\mu\n\a\b} - 4 \, R^{2}_{\mu\nu} + R^{2} \right] \phi\;
,
\end{equation}
so that the type A anomaly is indeed the Euler density,
\begin{equation}
{\cal A} = E_4 \; .
\end{equation}
As we mentioned, it is the unique scalar density whose
integral vanishes.  [Although there must be some ``descent identity"
which will make this result as obvious formally without explicit
calculation as it is physically by
scale-invariance, we have not been able to
find one.]  From the structure of the corresponding effective action, we
identify the $\Box^{-1}$ factor, which generates a $\d
(q^{2})$ discontinuity in the corresponding invariant amplitude; this
shows that the type A--producing mechanism is the same as for
chiral anomalies \cite{011,013}.

We now turn to the type B anomaly, originally found in \cite{001}.  It
first occurs at $d=4$, as there is no relevant invariant at $d=2$.
Consider the leading order Weyl invariant\footnote{Note, incidentally,
that the apparently equally acceptable form
$
\tilde{W}^0 = \int \, d^4x \, \sqrt{-g} \: C^{2} \, R/ \Box \; ,
$
which also yields ${\cal A} \sim \sqrt{-g}\, C^2$ is
\underline{not} permitted.  The procedure \cite{003} for
obtaining this ``integration" of the anomaly is invalid because the
required gauge choice suffers from the ambiguity
that different metrics, related by constant Weyl transformations, are
mapped to
the same final metric; equivalently,
$\tilde{W}$ cannot originate from
any leading order conformally invariant expression in $d = 4+2\e$:
while
say $(\Box + R)^{\e}$ is permitted in (21), $R^{\e}$ is not, being
nonpolynomial.} $W^0$,
\begin{equation}
W^0 = \e^{-1} \int \, d^dx \, \sqrt{-g} \: C \, \Box^{\e } \,
C \sim \int d^dx \, \sqrt{-g} \: C \, {\rm log} \Box \, C + O (\e ) \;
,
\;\;\;\;\; 2\e = d-4 \; .
\end{equation}
The all-orders completion of $W^0$, denoted by $W^1$, is ultraviolet
finite at $\e \rightarrow 0$, and hence will not contribute to the
anomaly (we have in fact explicitly constructed the first few terms of
$W^1$).  The counterterm needed to cancel the divergence in (21) is
\begin{equation}
W^c = \e^{-1} \int \, d^dx \, \sqrt{-g} \: C^{2} \m^{2\e} \; .
\end{equation}
The total action,
\begin{equation}
W = W^0 + W^1 - W^c
\end{equation}
has a finite limit, but $W^c$ explicitly breaks the conformal invariance
of $W^0 + W^1$.  Since $W$ is finite, we can interchange the order of
operations when calculating the anomaly:
\begin{equation}
{\cal A} \equiv \d_\phi (\lim W) =
\lim \d_\phi \, W =
-\lim \d_\phi W^c = -\sqrt{-g} \: C^2 \; .
\end{equation}
Alternately, we note that $W^0 - W^c$ has a term $\sim \log \Box
/\m^{2}$, and the explicit scale dependence thereby introduced again
yields (24), in accord with (4).  Given (24), we could also reconstruct
$W$,
since it must originate from our expression leading to the log
$\Box /\m^{2}$ term in the $d \rightarrow 4$ limit.

The above procedure also applies to other (nonchiral) anomalies in
systems, combining Weyl invariance with other symmetries, where a
scale must be introduced.  For example, the anomaly in the correlator
of the stress tensor and two vector currents at $d=4$ \cite{014} will be
encoded in an effective action involving both the external metric and
Maxwell field strength in $d=4+2\e$.  Following the pattern of (22--24),
one obtains this anomaly, proportional now to the Maxwell Lagrangian
density and derivable from an effective action term \cite{001} involving
$\log \Box$.  Indeed, whenever scale invariance is explicitly broken,
the anomalies will be proportional to the dimension four operators in
the theory.  [Of course, interactions will, as is well-known \cite{015},
complete the
coefficient to be the full beta function.]  As another instance, if
$D=4$ Weyl gravity were consistently quantizable, its anomaly would
just be the ``beta function" times (24).  In contrast, scale-preserving
anomalies have no particular relation to the beta function.

Let us now summarize the general structure of all the
anomalies in
any even dimension $d=2n$:

\begin{itemize}
\item
[a)] There is one type A anomaly, proportional to the Euler
density $E_{2n}$; its origin in the effective action is the (unique) term
containing $(n+1)$ Weyl tensors that vanishes in $d
\leq 2n + 1$.
\item
[b)] There is a rising number of type B anomalies,
corresponding to all the (non-vanishing) Weyl--invariant scalar
polynomials
each of whose terms is (symbolically) a product $\sim
R^{n-m}_{\m\n\a\b}
\Box^{m} (m < n-1)$
of $n$ curvature tensors and covariant Laplacians or suitably
contracted
derivatives (this is why
there is no type B in $d=2$ and only one in $d=4$).  The origin of each
of these
anomalies is the
counterterm in the effective action that cancels the logarithmic
divergence in the integral of the corresponding invariant
times
$\Box^{\frac{d-2n}{2}}$.
\end{itemize}

As a final illustration, consider $d=6$: The type A anomaly is $E_6$.
For type B,
appropriate polynomials are constructed from three curvature tensors
or two curvature tensors and one Laplacian (one tensor and two
Laplacians form a total derivative and hence
can be obtained as the variation of a local term).  There  are three
independent
Weyl-invariant
combinations, but only the (purely algebraic) first two are obvious:
$$
{\cal A}_1 = \sqrt{-g} \: C^{\mu\nu}_{\rho\s} \; C^{\rho\s}_{\a\b} \;
C^{\a\b}_{\m\n} \; ,
\eqno{(25{\rm a})}
$$
$$
{\cal A}_2 = \sqrt{-g} \: C^{\mu\nu}_{\rho\s} \; C^{\rho\a}_{\m\b}
\;
C^{\s\b}_{\n\a}\; ,
\eqno{(25{\rm b})}
$$
$$
{\cal A}_3 = \sqrt{-g} \left\{ C^{\mu\nu}_{\rho\s} \, \Box \,
C^{\rho\s}_{\m\n} +
2 \, C^{\m\n\rho\a} \: C_{\mu\nu\rho\b}R^\b_\a
- 3 \, C_{\mu\nu\rho\s}R^{\m\rho}R^{\n\s} \right.
$$
$$
{}~~~~~\left. - \textstyle{\frac{3}{2}} \: R^{\n\s} R_{\s\a} R^\a_\n +
\textstyle{\frac{27}{20}} \: R^{\mu\nu}R_{\mu\nu}R -
\textstyle{\frac{21}{100}}\: R^3 \right\} \; ;
\eqno{(25{\rm c})}
$$
the corresponding effective actions are the integrals of log $ \Box$
times these
terms.
The expressions (25) confirm an earlier analysis based on solving the
corresponding cohomology problem \cite{016}.

The complete classification of the Weyl anomaly structure proposed
here raises some interesting possibilities for the type A anomaly, since
it
is a true analogue of the chiral anomaly.  One might at first sight hope
that some of the
striking
features of the latter, such as the nonrenormalization theorem
\cite{017}, the relation to central extensions of the corresponding
algebra
\cite{018} and consistency conditions \cite{019}, could perhaps be
present here as well.
However, the explicit scale dependence introduced by renormalization
of the matter systems
(and the consequent possible mixing of the two types)
once interactions are included
poses a major obstacle to such hopes in realistic situations.
In this context, we remark that while the
Zamolodchikov theorem \cite{006} in $d=2$ is related to the type A
anomaly, its possible generalization to $d=4$ proposed in
\cite{008}
involves type B.

\noindent{\bf Acknowledgements}

This work was supported by the US--Israel Binational Science
Foundation grant 89--00140, and by NSF grant PHY88--04561.
Very useful discussions with L. Bonora, S. Elitzur, R. Iengo, and E.
Rabinovici are gratefully acknowledged.
We thank A. Barvinsky, S. Fulling, L. Parker, P. van
Nieuwenhuizen,  and A. Vilkovisky for
interesting conversations and correspondence.

\end{document}